\documentclass[a4paper,fleqn,usenatbib]{mnras}

\usepackage[T1]{fontenc}
\usepackage{ae,aecompl}

\usepackage{graphicx}	
\usepackage{amsmath}	
\usepackage{amssymb}	
\usepackage{multirow}

\def\vsini{$v\,\sin\,i$}              
\def\vsini{\hbox{$v$\,sin\,$i$}}      

\title[Close-in, massive planets of A-type stars]{Lack of close-in, massive planets of main-sequence A-type stars from Kepler} 

\author[Silvia Sabotta et al.]{
Silvia Sabotta$^{1}$\thanks{E-mail: sabotta@tls-tautenburg.de},
Petr Kabath$^{2}$,
Judith Korth$^{3}$, 
Eike W. Guenther$^{1}$, \newauthor
Daniel Dupkala$^{2}$, 
Sascha Grziwa$^{3}$,
Tereza Klocova$^{2}$ and
Marek Skarka$^{2,4}$
\\
$^{1}$ Th\"uringer Landessternwarte Tautenburg, Sternwarte 5, 07778
Tautenburg, Germany\\
$^{2}$ Astronomical Institute, Czech Academy of Sciences, Fri\v{c}ova 298, 25165, Ond\v{r}ejov, Czech Republic\\
$^{3}$ Rhenish Institute for Environmental Research, University of Cologne, Cologne, Germany\\
$^{4}$ Department of Theoretical Physics and Astrophysics, Masaryk University, Kotl\'{a}\v{r}sk\'{a} 2, 61137 Brno, Czech Republic}

\date{Accepted 2019 August 2. Received 2019 July 27; in original form 2019 April 17}

\pubyear{2019}

\begin{document}
\label{firstpage}
\pagerange{\pageref{firstpage}--\pageref{lastpage}}
\maketitle

\begin{abstract}
Some theories of planet formation and evolution predict that intermediate-mass stars host more hot Jupiters than Sun-like stars, others reach the conclusion that such objects are very rare.  By determining the frequencies of those planets we can test those theories.

Based on the analysis of \emph{Kepler} light curves it has been suggested that about 8 per cent of the intermediate-mass stars could have a close-in substellar companion. This would indicate a very high frequency of such objects. Up to now, there was no satisfactory proof or test of this hypothesis.

We studied a previously reported sample of 166 planet candidates around main-sequence A-type stars in the \emph{Kepler} field. We selected six of them for which we obtained extensive long-term radial velocity measurements with the Alfred-Jensch 2-m telescope in Tautenburg and the Perek 2-m telescope in Ond\v{r}ejov. We derive upper limits of the masses of the planet candidates. We show that we are able to detect this kind of planet with our telescopes and their instrumentation using the example of MASCARA-1\,b. 

With the transit finding pipeline EXOTRANS we confirm that there is no single transit event from a Jupiter-like planet in the light curves of those 166 stars. We furthermore determine that the upper limit for the occurrence rate of close-in, massive planets for A-type stars in the \emph{Kepler} sample is around 0.75 per cent.

We argue that there is currently little evidence for a very high frequency of close-in, massive planets of intermediate-mass stars.
\end{abstract}




\section{Introduction}

\label{sectI}

While many studies have been carried out to determine the frequency of planets of Sun-like stars, there is a considerable lack of information about the frequency of close-in planets of intermediate mass stars (IMSs) in the range $1.3 M_{\odot} \leq M_{\odot} \leq 3.2 M_{\odot}$.  This is very unfortunate, because the various theories of planet formation make different predictions about the frequencies of planets orbiting IMSs, particularly for massive planets at short orbital periods.  We can thus test the theories of planet formation by determining the frequency of planets orbiting IMSs particularly if we determine the frequency of planets in short orbital periods. Most theories predict that the frequency of massive planets should increase with the mass of the star \citep{laughlin1993, ida2005, kennedy2008, alibert2011, mordasini2012, hasegawa2013}. Some models however predict the opposite \citep{kornet2006, boss2005}.

Direct-imaging surveys of main-sequence A-type stars have revealed a number of planets e.g. \cite{marois2008} or \cite{marois2010}. \cite{vigan2012} analysed the frequency of planets of main-sequence IMSs based on direct imaging surveys statistically. They conclude that stars more massive than the sun have a higher frequency of massive planets, at least at semi-major axis ranges between 10--300\,au. 

Unfortunately, classical radial-velocity (RV) surveys are not very suitable for studying A-type main-sequence stars. They have a relatively small number of spectral lines and rotate rapidly. 

One way out of this dilemma is to observe post-main sequence stars, the so-called retired A-type stars. RV surveys of retired A-type stars are very successful and have resulted in many discoveries e.g. \cite{johnson2010a, johnson2010b}, \cite{lovis2007}. The statistical analysis by \cite{johnson2010a, johnson2010b} indicates a higher frequency of massive planets for intermediate mass stars than for Sun-like stars. However, the results of \cite{johnson2010a,johnson2010b} have been criticized by \cite{lloyd2011,lloyd2013} who argued that the mass determination of post-main sequence stars through spectroscopy and evolutionary tracks is not reliable. Consequently the masses have been reevaluated with asteroseismology \citep{north2017, stello2017} and for a subsample of retired A-stars in binary star systems \citep{ghezzi2015}. The so-determined masses are in agreement or 15--20\,per cent lower than the ones originally derived by \cite{johnson2010a, johnson2010b}. \cite{ghezzi2018} reanalysed the masses of 245 subgiants spectrosopically and reached to the same conclusion. It therefore appears that the frequency of massive planets of IMSs is higher than that of Sun-like stars.

However, these surveys have only detected planets at distances larger than 0.5\,au from the host star. A more conclusive test would be to study the frequency of short-period (less than 10\,d), massive planets for which theories make very different predictions. \cite{hasegawa2013} predict that the frequency of hot Jupiters increases dramatically with the mass of the host star.  The reason for this is that hot Jupiters form in a 'dead-zone' close to the star, which contains a large amount of matter. In sharp contrast to this, \cite{stephan2018} call A-type stars 'the destroyers of worlds'. Most A-type stars have stellar binary companions that can strongly affect the dynamical evolution of planets around either star through the eccentric Kozai-Lidov mechanism \citep[e.g.][]{naoz2016,naoz2012, petrovich2015, anderson2016}. The binary fraction of A-type stars is much higher than of Sun-like stars \citep[e.g. 84$\pm$ 11\,per cent in][]{moe2017}.  \cite{stephan2018} predict that only 0.15\,per cent of A-type stars will host hot Jupiters during their main-sequence lifetimes. There are thus completely contradicting predictions from theory, the rate of hot Jupiters could be higher than for G-stars, e.g., $\geq 1.2\pm0.38$ per cent \citep{wright2012}, or the rate could be as low as 0.15\,per cent.  Determining the frequency of hot Jupiters orbiting A-type stars is an excellent test of the theories of planet formation and evolution.

Since classical RV-surveys are not suitable to effectively detect planets around A-type stars, a better strategy is to use transit surveys. A number of transiting hot Jupiters of A-type main-sequence stars have been found. The first one was WASP-33\,b/HD 15082\,b \citep{cameron2010} which has an orbital period of only 1.2 days. Using 248 RV-measurements obtained with the Alfred-Jensch telescope we determined its mass to $2.1\,M_\text{Jup}\pm0.2\,M_\text{Jup}$ \citep{lehmann2015}. 

Other hot Jupiters of A-type stars found in transit surveys are Kepler-13\,b \citep{shporer2011}, HAT-P-57\,b \citep{hartman2015}, KELT-17\,b \citep{zhou2016}, MASCARA-1\,b \citep{talens2017}, MASCARA-2\,b/KELT-20\,b, HD\,185603\,b \citep{talens2018, lund2017}, WASP-189\,b/HR5599\,b \citep{anderson2018}, KELT-21\,b/HD\,332124\,b \citep{johnson2018}, KELT-9\,b/HD\,195689\,b \citep{gaudi2017}, KELT-19A\,b \citep{siverd2017} and MASCARA-4\,b/bRing-1\,b \citep{dorval2019}. Studying the hot Jupiter population around hot stars is interesting not only because their frequency constrains the theory of planet formation, but also because they are ideal laboratories to study atmospheric escape and the properties of planet atmospheres. 

\begin{figure}
\centering
\includegraphics[width=\linewidth]{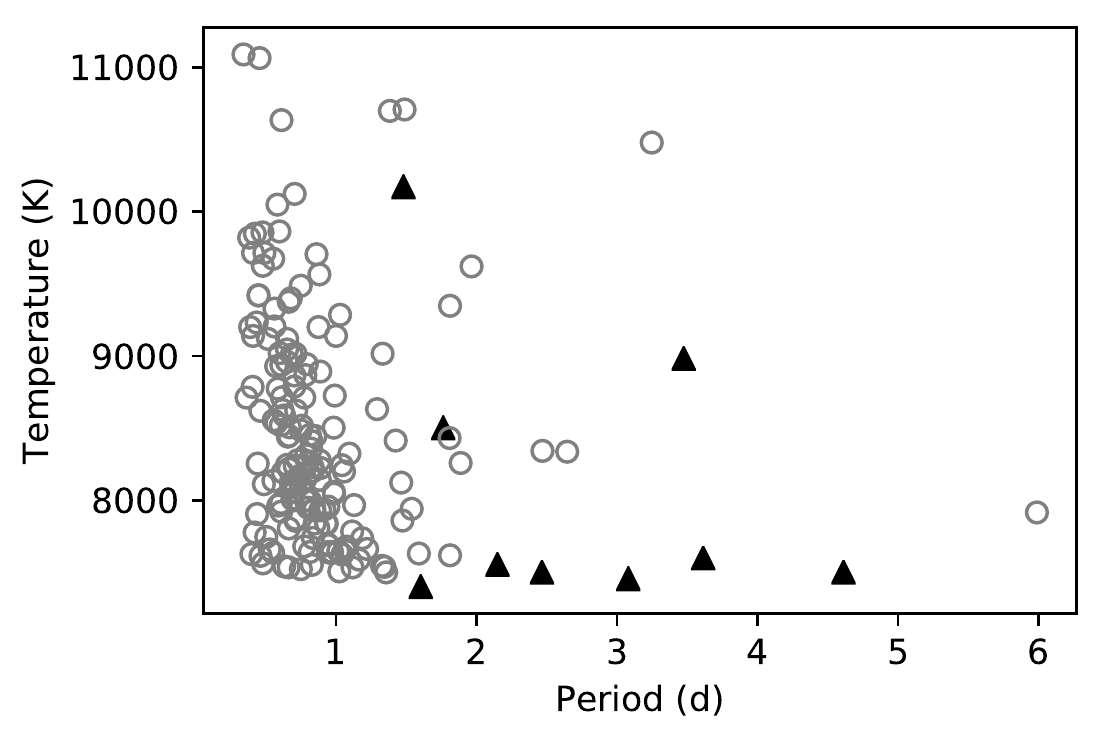}
\caption{Black triangles: detected and confirmed planets of A-type stars. Gray circles: additional possible planets according to Balona 2014 (see \emph{exoplanet.eu}).}
\label{planet_number}
\end{figure}

\cite{balona2014} found peculiar features in the periodograms of the light curves of 166 A-type stars. Given that he found this feature in 166 out of 1974 A-type stars, he concludes that about 8 per cent of the A-type stars could have massive planet or brown dwarf companions which have orbital periods of 6 days or less. If this is true, the number of known planets around A-type stars would increase drastically (see Fig.~\ref{planet_number}). 

Although this is an exciting result, the hypothesis has not been tested yet. It is thus time to shed new light onto the question, if the fraction of A-type stars that have a close-in massive planet is higher, or of it is much lower than for Sun-like stars.

In this paper we will present our investigation of the 166 A-type stars reported by \cite{balona2014}. Our paper presents radial velocity data obtained by the echelle spectrographs at the Alfred Jensch and the Perek telescope. The datasets and results are described in Section~\ref{sectII}. Furthermore we looked for transits in the sample. We then investigated the hot Jupiter frequency in the Kepler A-star sample. The analysis is shown the Sections~\ref{sectIV} and \ref{sectV}. Discussion and our conclusions can be found in Sections~\ref{sectVI} and \ref{sectVII}

\section{RV-measurements}
\label{sectII}

\cite{balona2014} analysed the light curves of 1974 A-type stars obtained with the \emph{Kepler} satellite \citep{borucki2010}. They found peculiar features in the periodograms of 166 of them. The peculiar feature consists of a broad peak (possibly due to differential rotation) and a sharp feature at a slightly higher frequency. \cite{balona2014} claim that the sharp feature could be caused by a planet with a similar orbital period as the rotational period.

Since \cite{balona2014} interpret the peculiar feature as a result of the beaming and ellipsoidal modulations \citep{mazeh2010}, the object causing this would have to be a massive hot Jupiter, or a brown dwarf, or even low-mass stellar companion \citep{talor2015}. This hypothesis can thus be tested by obtaining RV-measurements with meter-class telescopes.  However, A-type stars are usually rapidly rotating and have only few spectral lines. The expected RV-error is proportional to the \vsini~of the star. Therefore, the expected RV precision is 20-60 times lower than for a comparable G-type star \citep{hatzes2016}. One way to deal with the lower RV precision is to take a lot of measurements. For example, we required 248 RV-measurements to determine the mass of WASP-33b \citep{lehmann2015}, which has a K-amplitude of around 300\,m\,s$^{-1}$. 

By selecting a number of stars from this list and obtaining several tens of RV measurements of each one, we can put first constrains on the masses of potential companions. Therefore, we randomly selected a small subsample of six stars that were observed with the 2-m telescopes in Tautenburg and Ond\v{r}ejov. An overview of the properties of this subsample is provided in Table~\ref{balonastars}.  

\begin{table}
\caption{The sample}
\begin{tabular}{l l l l l l l}
\hline
\noalign{\smallskip}
name            & $\alpha$ (J2000.0)$^1$ & SpT$^2$ & $m_\text{v}^3$ & parallax$^1$ \\
other name      & $\delta$ (J2000.0)$^1$ & ~ & ~ &  (mas) \\
\hline
KIC\,3766112 & 19$^{\text{h}}$44$^{\text{m}}$1.9$^{\text{s}}$ & \multirow{2}{*}{A0} & \multirow{2}{*}{11.3} & \multirow{2}{*}{$0.92\pm0.04$} \\
HD\,225570  &  +38$^\circ$52$'$58.5$''$\\[0.2cm]
KIC\,4944828 & 19$^{\text{h}}$47$^{\text{m}}$47.2$^{\text{s}}$  & \multirow{2}{*}{A5}  & \multirow{2}{*}{9.9}  & \multirow{2}{*}{$2.04\pm0.03$} \\
HD\,225856  &  +40$^\circ$0$'$57.3$''$ \\[0.2cm]
KIC\,7352016 & 19$^{\text{h}}$13$^{\text{m}}$14.2$^{\text{s}}$  & \multirow{2}{*}{A9} & \multirow{2}{*}{12.0} & \multirow{2}{*}{$1.15\pm0.02$ }\\
TYC\,312925431 & +42$^\circ$54$'$49.5$''$  \\[0.2cm]
KIC\,7777435 & 19$^{\text{h}}$55$^{\text{m}}$24.4$^{\text{s}}$  & \multirow{2}{*}{A2} & \multirow{2}{*}{10.7} & \multirow{2}{*}{$1.39\pm0.03$} \\
HD\,188874  & +43$^\circ$29$'$48.0$''$  \\[0.2cm]
KIC\,9222948 & 19$^{\text{h}}$34$^{\text{m}}$46.7$^{\text{s}}$ & \multirow{2}{*}{A1} & \multirow{2}{*}{10.2} & \multirow{2}{*}{$2.25\pm0.05$} \\
BD+45\,2925 & +45$^\circ$37$'$11.8$''$   \\[0.2cm]
KIC\,9453452 & 19$^{\text{h}}$4$^{\text{m}}$36.1$^{\text{s}}$ & \multirow{2}{*}{A4} & \multirow{2}{*}{10.6} & \multirow{2}{*}{$1.57\pm0.03$ }\\
TYC\,354130011 & +46$^\circ$3$'$37.9$''$ \\
\hline
\end{tabular}
\label{balonastars}
\\
$^1$ Taken from Gaia DR2 \citep{prusti2016,brown2018} \\
$^2$ Taken from \cite{frasca2016} \\
$^3$ Taken from \cite{hog2000} \\
\end{table}

\subsection{Data obtained with the Alfred-Jensch 2-m telescope at Tautenburg observatory}\label{sec:Taut}

The sample of six A-type star planet candidates was monitored with the 2-m Alfred Jensch telescope of the Th\"uringer
Landessternwarte Tautenburg which is equipped with an echelle spectrograph with resolving power of $\lambda / \Delta\lambda =35000$ with the two arsec slit used. We observed MASCARA-1\,b as a reference object, because it is an A-type star with a known planet.

The typical signal to noise of the obtained spectra is listed in Table~\ref{ston1}:

\begin{table}
\caption{Typical S/N of the Tautenburg observations}
\centering
\begin{tabular}{llll}
  \hline
star & S/N & star & S/N   \\
\hline
KIC\,3766112 & 25 -- 45 & KIC\,7777435 & 30 -- 60 \\
KIC\,4944828 & 40 -- 80 & KIC\,9222948 & 40 -- 75 \\
KIC\,7352016 & 30 -- 35 & KIC\,9453452 & 30 -- 60 \\
MASCARA-1  & 80 -- 120 & ~ & ~ \\
\hline
\end{tabular}
\label{ston1}
\end{table}

The data-reduction followed the usual steps, bias-subtraction, flat-fielding, removal of Cosmic Rays, scattered light subtraction, extraction, wavelength calibration and normalization. All methods were combined using the Tautenburg Spectroscopy Pipeline -- $\tau$-spline. The pipeline makes use of standard \small{IRAF}\footnote{IRAF is distributed by the National Optical Astronomy Observatories, which are operated by the Association of Universities for Research in Astronomy, Inc., under cooperative agreement with the National Science Foundation.} and PyRaf routines\footnote{PyRaf is a product of the Space Telescope Science Institute, which is operated by AURA for NASA} and the Cosmic Ray code by Malte Tewes based on the method by \cite{vandokkum2001}.

We used the telluric lines in order to account for instrumental shifts. The telluric shift was obtained by cross correlation of a telluric O$_2$-template (extracted with the ESO Program Molecfit, see \cite{smette2015} and \cite{kausch2015}) with the object spectra. 

\subsection{Data obtained with the Perek 2-m telescope at Ond\v{r}ejov observatory}

The advantage of a monitoring network of telescopes at central Europe is that we can obtain a better coverage of the data sets. Therefore, we also used Perek 2-m telescope located at the Astronomical Institute of the Czech Academy of Sciences at Czech Republic. It is equipped with an Echelle Spectrograph (OES).

OES has resolving power $\lambda/\Delta\lambda = 44000$ and a slit width of 0.6\,mm, corresponding to 2 seconds of arc on the sky. Further details and description of instrument capability can be found in \citep{kabath2019}.

We monitored the same sample of stars as Alfred-Jensch telescope as described in Tab \ref{balonastars}. The data was reduced in the same way as described in Section~\ref{sec:Taut} for the Tautenburg data. To obtain a higher signal to noise 2-3 spectra were combined.

The typical signal to noise of the obtained spectra is listed in Table~\ref{ston2}:

\begin{table}
\caption{Typical S/N of the Ond\v{r}ejov observations}
\centering
\begin{tabular}{llll}
  \hline
star & S/N & star & S/N   \\
\hline
KIC\,4944828 & 20 -- 30 & KIC\,9222948 & 20 -- 40 \\
KIC\,9453452 & 15 & MASCARA-1  & 35 \\
\hline
\end{tabular}
\label{ston2}
\end{table}

\subsection{Results of the RV-measurements}
\label{sectIII}

The radial velocities were obtained by cross correlation. The template was obtained by co-adding all spectra of the star similar to the method described in \cite{anglada2012}. We obtained the RV error for each spectrum by cross correlating every order separately. After removing outliers we use the mean value as radial velocity and the standard deviation as the corresponding error.

The resulting RV values were analysed with the Radial Velocity Modeling Toolkit -- RadVel -- by \cite{fulton2018}. This algorithm uses a simple maximum-likelihood fit as an initial guess and afterwards performs a Markov-Chain Monte Carlo (MCMC) exploration to obtain the corresponding errors. The starting values for the fit are the periods published by \cite{balona2014} and we assume zero eccentricity as the planets are supposedly close-in.

\begin{figure}
\includegraphics[width=\linewidth]{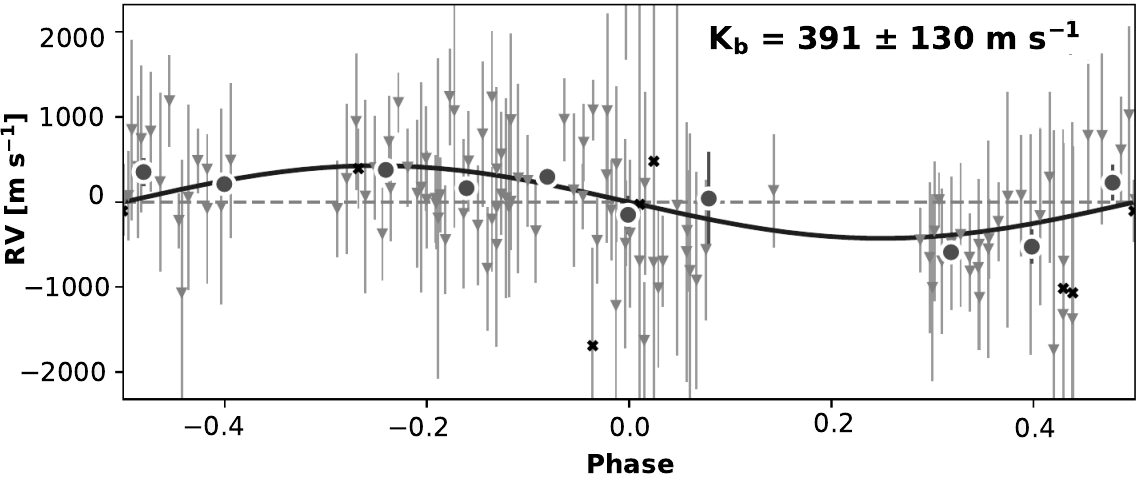}
 \caption{RV curve for MASCARA-1\,b. X: the Ond\v{r}ejov values; Triangles: the Tautenburg values; Circles: Binned radial velocities}
 \label{mascara}
\end{figure}

We have observed MASCARA-1\,b as a reference object to make sure that with our method we are capable of measuring masses of hot Jupiters around A-type stars. We used exactly the same method as for the other stars. We took 114 measurements in Tautenburg and analysed them in the same way as described above. We obtain a K-amplitude of 391,m\,s$^{-1}$ $\pm$ 130\,m\,s$^{-1}$ (see Fig.~\ref{mascara}) which is consistent with the K-amplitude reported by \cite{talens2017} of 400\,m\,s$^{-1}$ $\pm$ 100\,m\,s$^{-1}$.

As we do not have as many observations for the 'Balona stars', we will only obtain upper limits for the possible planet masses. To obtain the upper limits we use the fitting algorithm of RadVel with a high initial K-amplitude value. The K-amplitude of the initial fit can be used as an estimate of the upper limit similar to what \citep{talens2018} did.

\begin{figure}
\includegraphics[width=\linewidth]{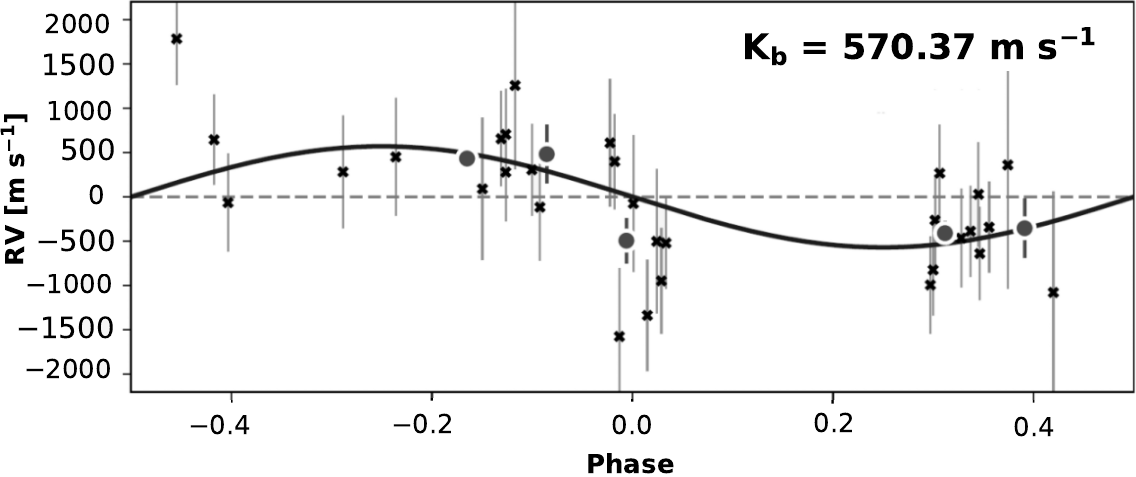}
 \caption{Example of upper limit for MASCARA-1\,b with only 30 data points from Tautenburg, black curve is the maximum-likelihood fit}
 \label{mascara_short}
\end{figure}

To show how this works we reduce the number of measurements of MASCARA-1\,b randomly to 30 which is in the order of magnitude of the measurements we obtained for the 'Balona stars'. Still we can fit a radial velocity curve to the data. The result is shown in Fig.~\ref{mascara_short}. The K-amplitude in this example is 570\,m\,s$^{-1}$ which matches the upper value obtained from the MCMC algorithm we ran on the whole set of measurements. Running RadVel's MCMC algorithm now gives zero as the most probable K-amplitude.

\begin{figure}
\includegraphics[width=\linewidth]{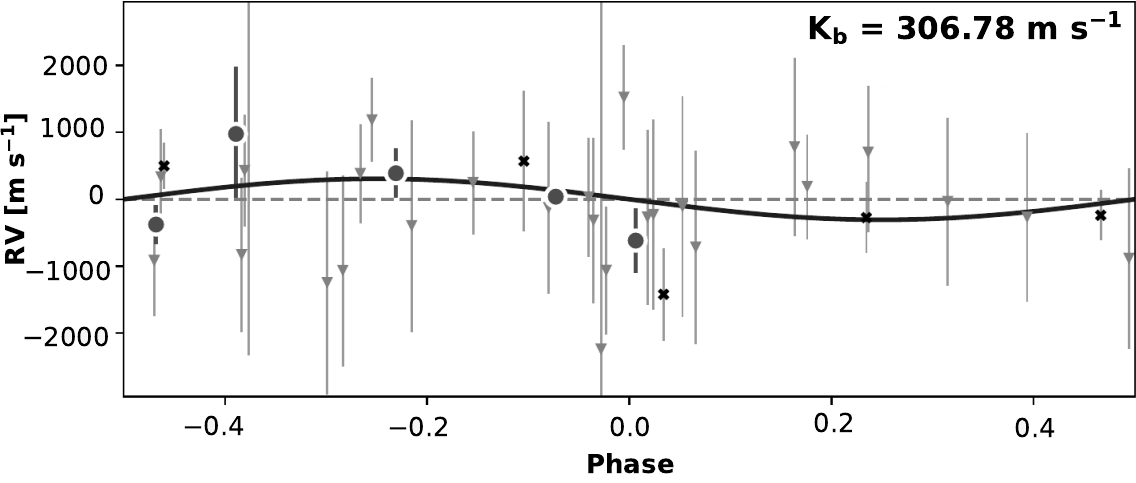}
\includegraphics[width=\linewidth]{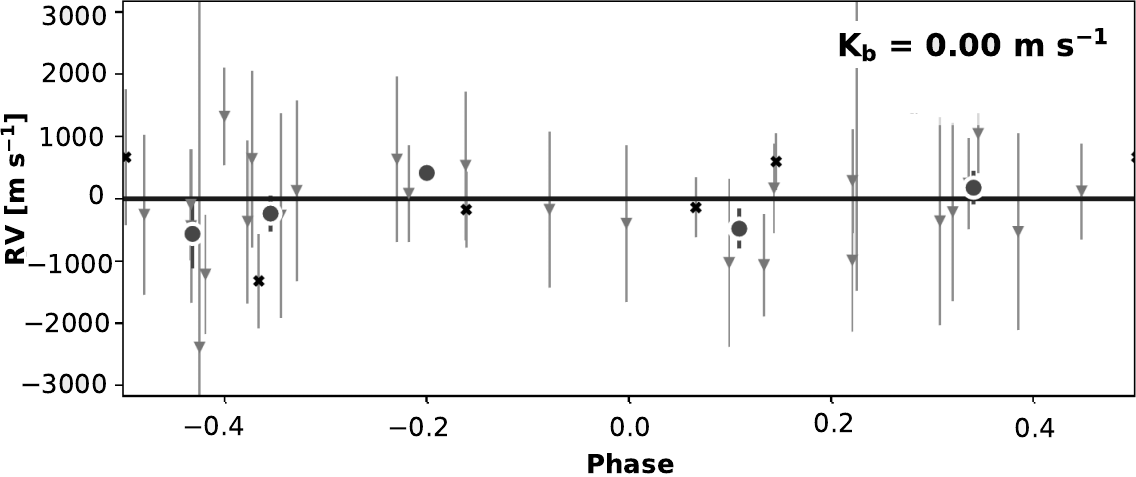}
 \caption{RV curve of KIC\,9222948; upper panel: simple maximum-loglikelihood fit to the data; lower panel: most probable K-amplitude after MCMC}
 \label{balona1}
\end{figure}

All six RV-curves of the 'Balona-stars' can be fitted with a simple maximum-likelihood fit as well. As soon as we run the MCMC algorithm the K-amplitudes turn out to be not significant (see e.g. Fig.~\ref{balona1}). We therefore use the K-amplitude of the initial fit and the known stellar masses to obtain upper limits for the possible companion masses with 1\,$\sigma$ confidence. The upper limits with 99 per cent confidence range from 3.8\,M$_\text{Jup}$-7.3M$_\text{Jup}$ and are summarised in Table~\ref{upperlimit}. The similar upper limits are not surprising as the variations are most probably only due to measurement errors. We obtained a similar amount of measurements for each star and a similar S/N for the individual spectra. Therefore instrumental effects would necessarily lead to similar results. All other radial velocity curves are reported in the appendix (Fig.~\ref{balona2}, \ref{balona3}, \ref{balona4}, \ref{balona5} and Fig.~\ref{balona6}).

\begin{table}
\centering
  \begin{tabular}{lllll}
  \hline
star & star mass & max. K & period & upper limit  \\
~ & (M$_\odot$) & (ms$^{-1}$) & (d) & (M$_\text{Jup}$) \\
\hline
KIC\,3766112 & 2.2 & 450 & 0.45 & 7.3 \\
KIC\,4944828 & 2.0 & 330 & 0.93 & 6.3 \\
KIC\,7352016 & 3.0 & 290 & 0.93 & 7.3 \\
KIC\,7777435 & 2.1 & 210 & 0.68 & 3.8 \\
KIC\,9222948 & 2.3 & 310 & 1.29 & 7.3 \\
KIC\,9453452 & 2.1 & 270 & 0.61 & 4.5 \\
\hline
\end{tabular}
    \caption{Summary of the upper limits.} 
    \label{upperlimit}
\end{table}

\section{Transit Analysis}
\label{sectIV}

The possible planets published by \cite{balona2014} should all have relatively short periods and large radii. This makes it obvious to look for transits. 

From the average stellar radius $R_\text{star}\approx$~2.3\,$R_\odot$ and the typical orbital radius $a\approx$~0.02\,au \citep{balona2014} we can calculate the transit probability $P_\text{transit} = \frac{R_\text{star}}{a}$ of every single 'possible planet'. It is around 53 per cent. Therefore, the probability that all 166 possible planets are not transiting is as low as 10$^{-55}$. \cite{balona2014} argues that the transits could be overlooked because the transit takes a big part of the orbit and because it could be hidden in the activity.

We therefore decided to model the transits in the light curves and see if we can retrieve them with the transit finding pipeline EXOTRANS \citep{grziwa2012,korth2019}. 

\begin{figure}
\centering
\includegraphics[width=\linewidth]{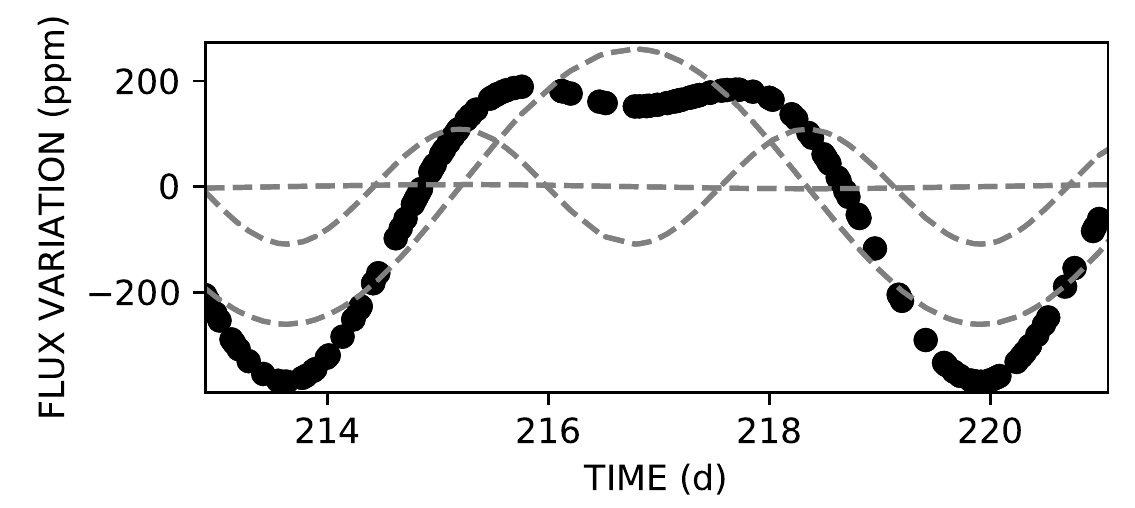}
 \caption{Model of ellipsoidal-, beaming- and reflection effect. Dashed lines: single effect, dotted line: combined effects}
   \label{model}
\includegraphics[width=0.9\linewidth]{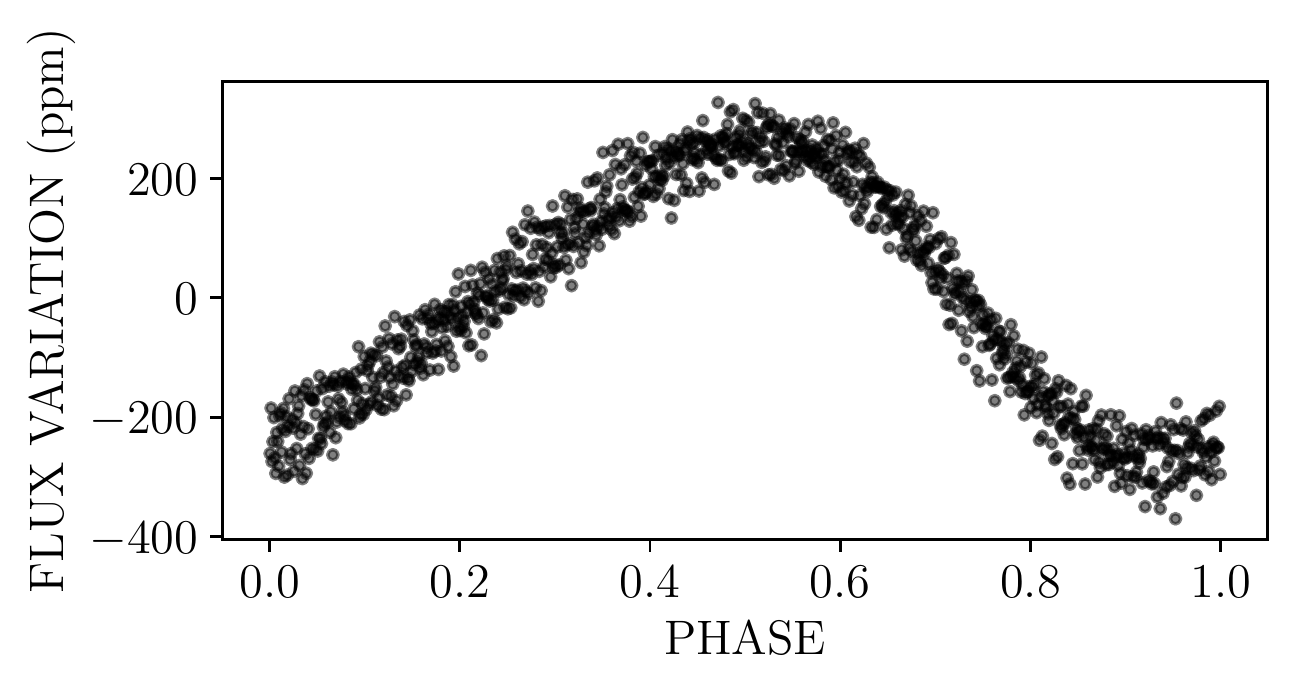}
 \caption{Light curve of KIC\,9222948, phase-folded with the period from \protect\cite{balona2014} and binned}
  \label{phasefolded}
\end{figure}

The possible planets need to have at least the mass and radius of Jupiter. This is because the ellipsoidal-, beaming- and reflection effects are mass and radius dependent \citep{mazeh2010,faigler2011}. In Fig.~\ref{model} we show a simple model of the three effects for a Jupiter-like planet. The figure shows that it is possible to retrieve the amplitude of the light curve variations due to a none transiting companion. In Fig.~\ref{phasefolded} we show one example of a phase-folded light curve of KIC\,9222948. From this we know that we cannot retrieve the shape of the light curve with the model. This could be due to an overlap with an effect on the star, e.g. spots, oscillations, pulsations.

We list the expected transit depths of the six stars we have analysed in Section~\ref{sectIII} for 1\,R$_\text{Jup}$ size planets in Table~\ref{transitdepth}.

\begin{figure}
 \centering
\begin{minipage}{\linewidth}
\includegraphics[width=\linewidth]{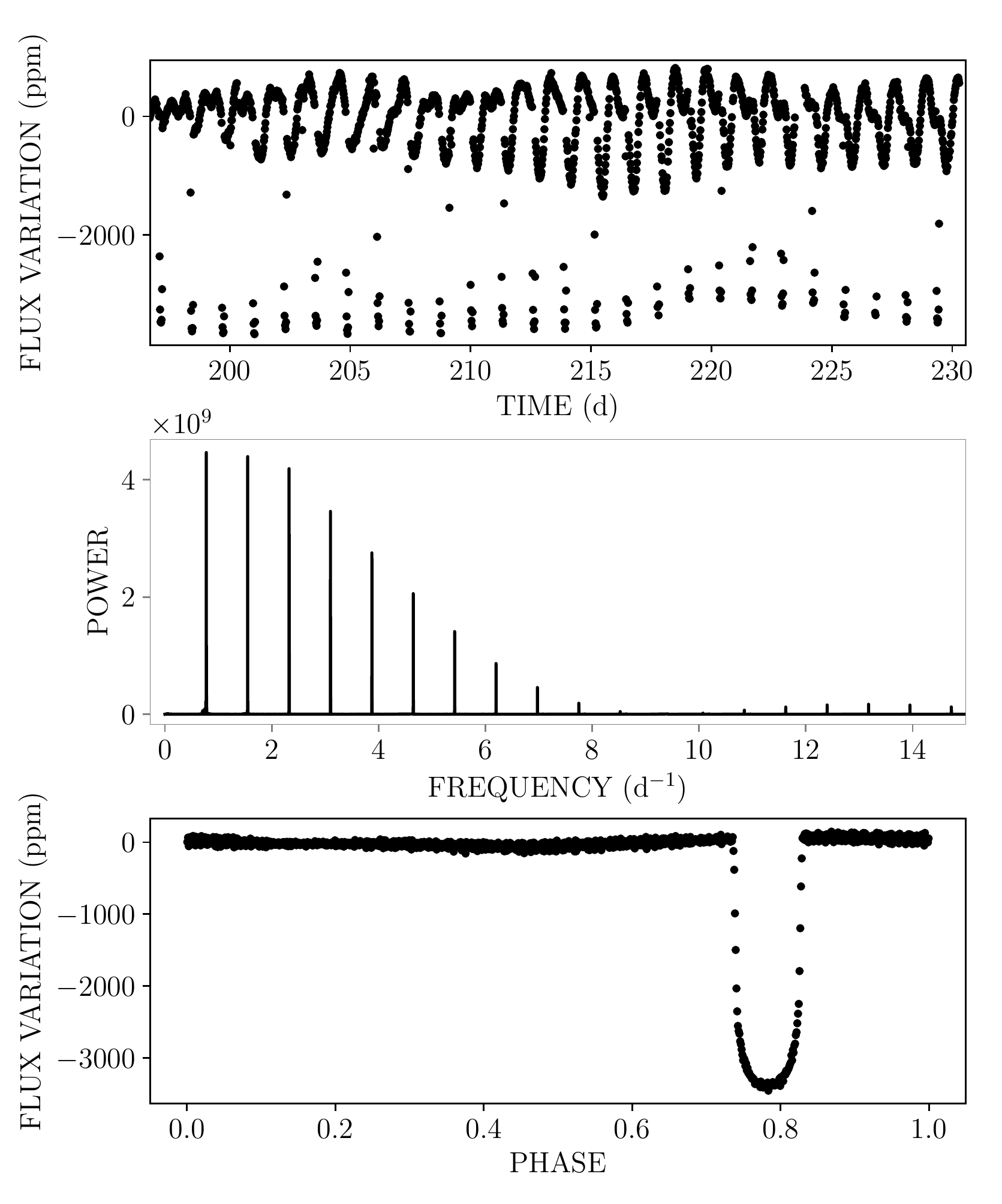}
\end{minipage}
\caption{Transit model of a Jupiter size object in \emph{Kepler} light curve of KIC\,9222948; top panel: 30\,d section of the time series, middle panel: periodogram of the light curve, lower panel: phase-folded and binned light curve}
\label{blindtest}
\end{figure}

To model the transit we used the actual light curves and combined them with a transit model. This model was obtained by using the code of \cite{parviainen2015} with the radii obtained from \cite{berger2018} and the limb darkening coefficients from \cite{sing2010}. In Fig.~\ref{blindtest}, we show one model transit of a Jupiter size object in a \emph{Kepler} light curve. All modeled transits can be easily retrieved by the detection software.

We consequently ran the transit finding algorithm on all 166 A-type stars and could not detect a single transit event from a Jupiter-like planet. All detected signals were much too small for a Jupiter like transit and were most likely activity signals.

\begin{table} \caption{Summary of the expected transit depth.} 
\centering
  \begin{tabular}{llll}
  \hline
star & Period (d) & R$_\text{star}$ ($R_\odot$) & Transit depth if \\
 ~ & ~ & ~ & R$_\text{p} \approx$ R$_\text{Jup}$ \\
\hline
KIC\,3766112 & 0.45\,d & 2.320$^1$ & 0.19\,\% \\
KIC\,4944828 & 0.93\,d & 2.457$^1$ & 0.17\,\% \\
KIC\,7352016 & 0.93\,d & 2.073$^1$ & 0.23\,\% \\
KIC\,7777435 & 0.68\,d & 2.572$^1$ & 0.15\,\% \\
KIC\,9222948 & 1.29\,d & 1.821$^1$ & 0.30\,\% \\
KIC\,9453452 & 0.61\,d & 2.166$^1$ & 0.21\,\% \\
\hline
\end{tabular}
    \label{transitdepth}\\
  $^1$  Taken from \cite{berger2018}
\end{table}

\section{Hot Jupiter Frequency with \emph{Kepler} Analysis}
\label{sectV}

The \emph{Kepler} satellite observed around 2000 A-type stars \citep{brown2011}. It found exactly one transiting hot Jupiter around A-type stars (Kepler-13\,A\,b). There are three other confirmed candidates: 
\begin{itemize}
 \item Kepler-340\,b with an orbital period of 22.9\,d and a size of 3.36\,R$_\oplus$,
 \item Kepler-340\,c with an orbital period of 14.8\,d and a size of 2.49\,R$_\oplus$,
 \item Kepler-1115\,b with an orbital period of 23.5\,d and a size of 1.7\,R$_\oplus$.
\end{itemize}
None of them can be classified as a hot Jupiter. The only other KOI around A-type stars with orbital periods less than 10\,d are KOI\,80, KOI\,971, KOI\,6068, KOI\,7733 and have planet sizes of 96\,R$_\oplus$, 134\,R$_\oplus$, 100\,R$_\oplus$ and 4.5\,R$_\oplus$ respectively. Therefore, none of these objects is likely to be classified as a hot Jupiter later. KIC\,11180361 or KOI\,971 is also in the sample of \cite{balona2013} and can be classified as a binary system with the output of EXOTRANS. KOI\,80 and KOI\,6068 are more likely to be binary stars as well, considering their transit depth.

Assuming a high occurrence rate of hot Jupiters of around 1.2 per cent, there should be a around 26 hot Jupiters in the whole sample. Even assuming that they are so close that the transit probability is around 50 per cent, there should be only around 13 transiting hot Jupiters in the sample. Taking the lower value of 0.15 per cent only three hot Jupiters should be in the sample of which one or two would be transiting. For a planet orbiting its star in a 10 days orbit instead of a 1 day orbit (about 0.09\,au), the transit probability drops as low as 11.5 per cent. In this case three or less transits should be present for the whole \emph{Kepler} sample. 

To obtain a more quantitative statement on the expected number of transits we conduct a small simulation. We take the actual stellar radii of the whole \emph{Kepler} A-star sample from \cite{berger2018} -- around 2000 stars. We assign each of them a planet with a certain probability (8.4 per cent for a planet frequency of 8.4 per cent). Then this planet gets a random semimajor axis between 0.01\,au and 0.1\,au. From this we calculate the transit probability $P_\text{transit} = \frac{R_\text{star}}{a}$. Then we randomly draw a 1 or a 0 with exactly this probability -- equal to throwing a coin that shows 'transit yes' or 'transit no'. Then we count all the transits. We repeat this process 7000 times. After all 7000 simulations are finished, we obtain a number of simulations that show a certain number of transits. This number we translate to a probability density by normalization.

\begin{figure}
\centering
\begin{minipage}{0.79\linewidth}
\includegraphics[width=\linewidth]{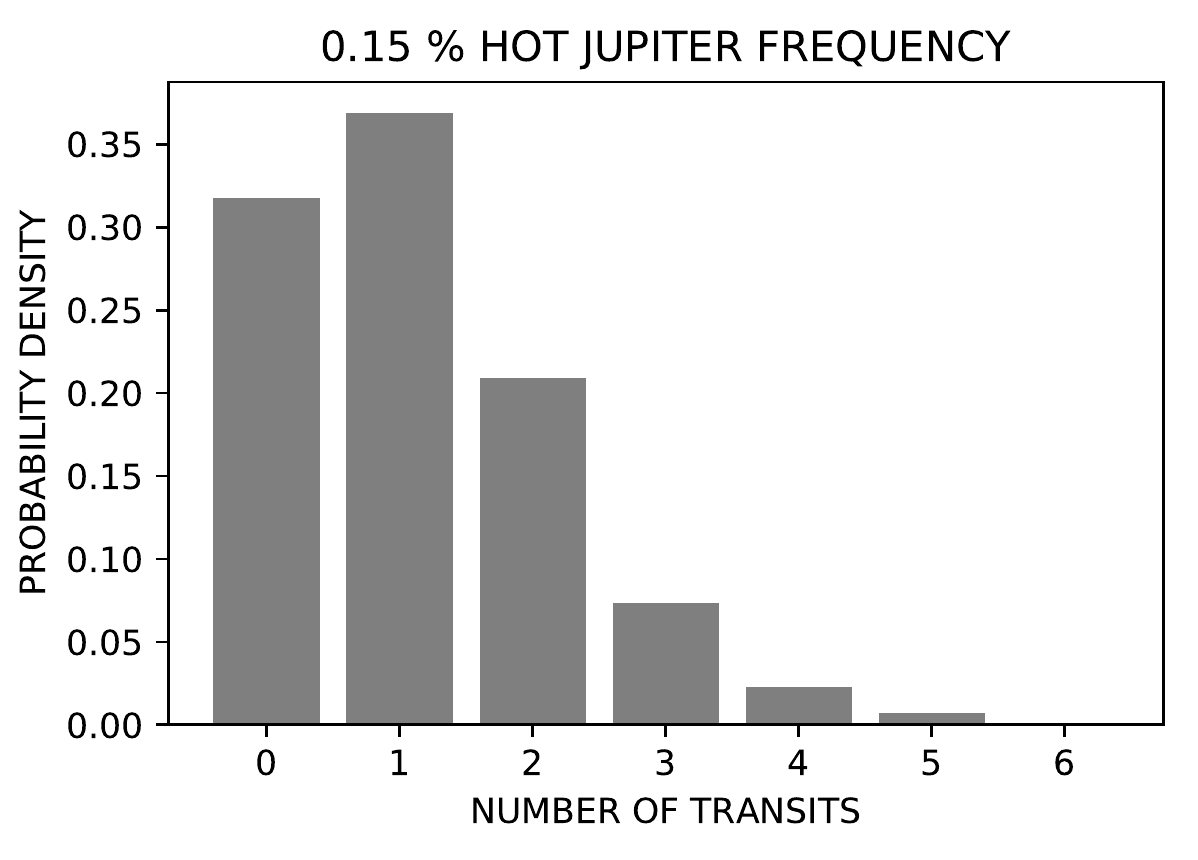}
\end{minipage}
\begin{minipage}{0.79\linewidth}
\includegraphics[width=\linewidth]{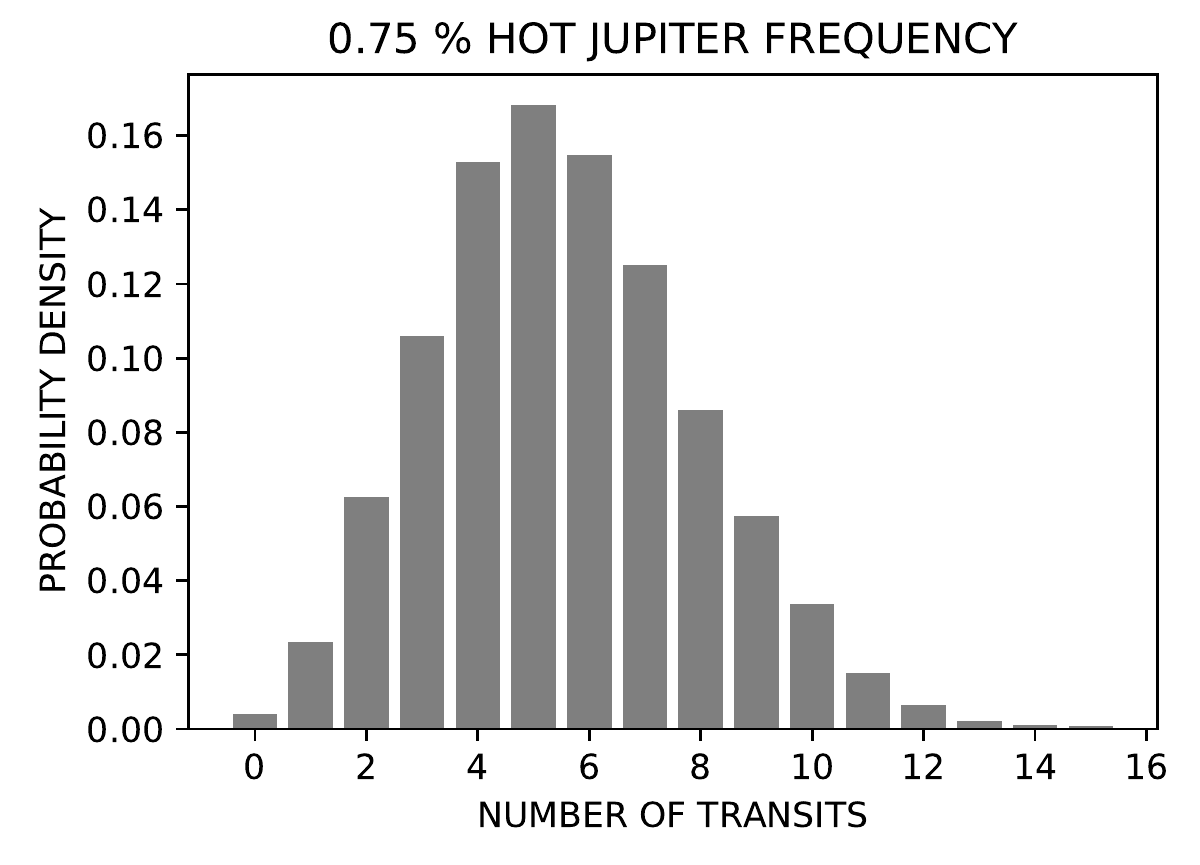}
\end{minipage}
\begin{minipage}{0.79\linewidth}
\includegraphics[width=\linewidth]{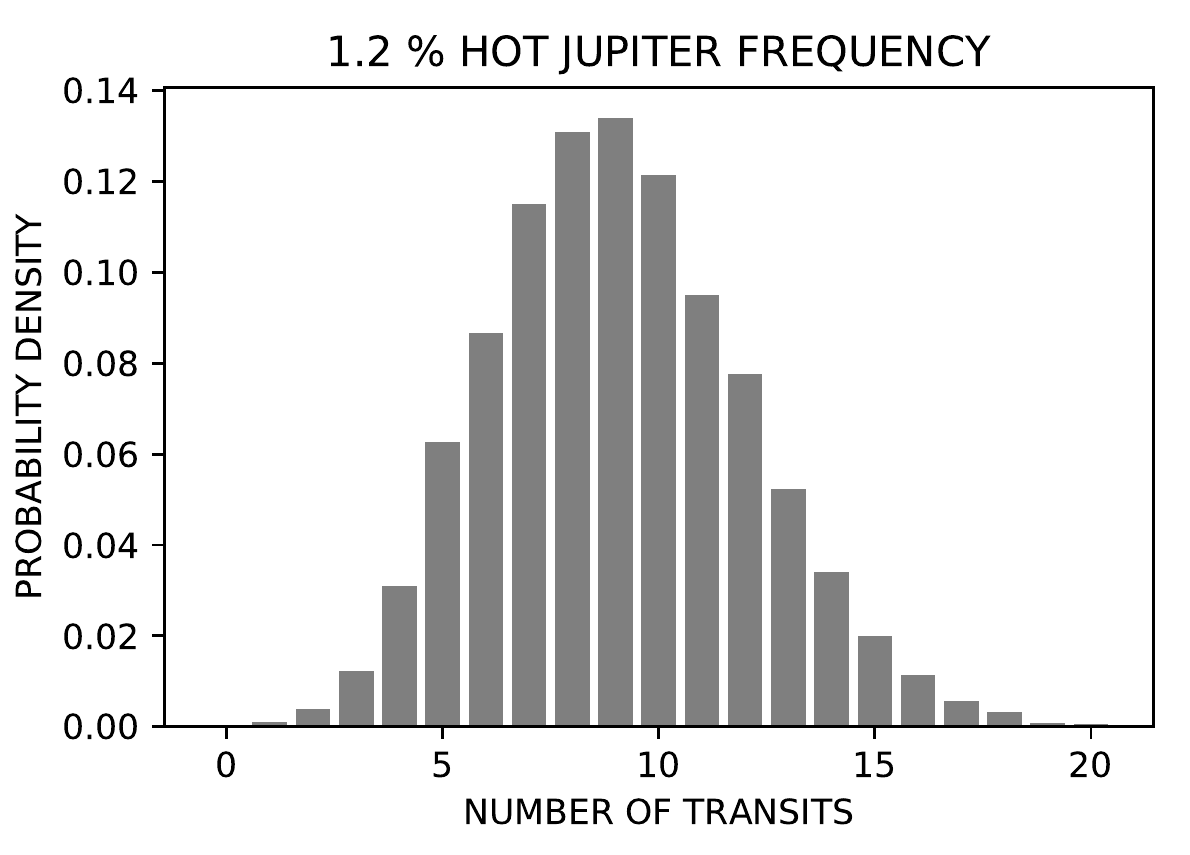}
\end{minipage}
\begin{minipage}{0.79\linewidth}
\includegraphics[width=\linewidth]{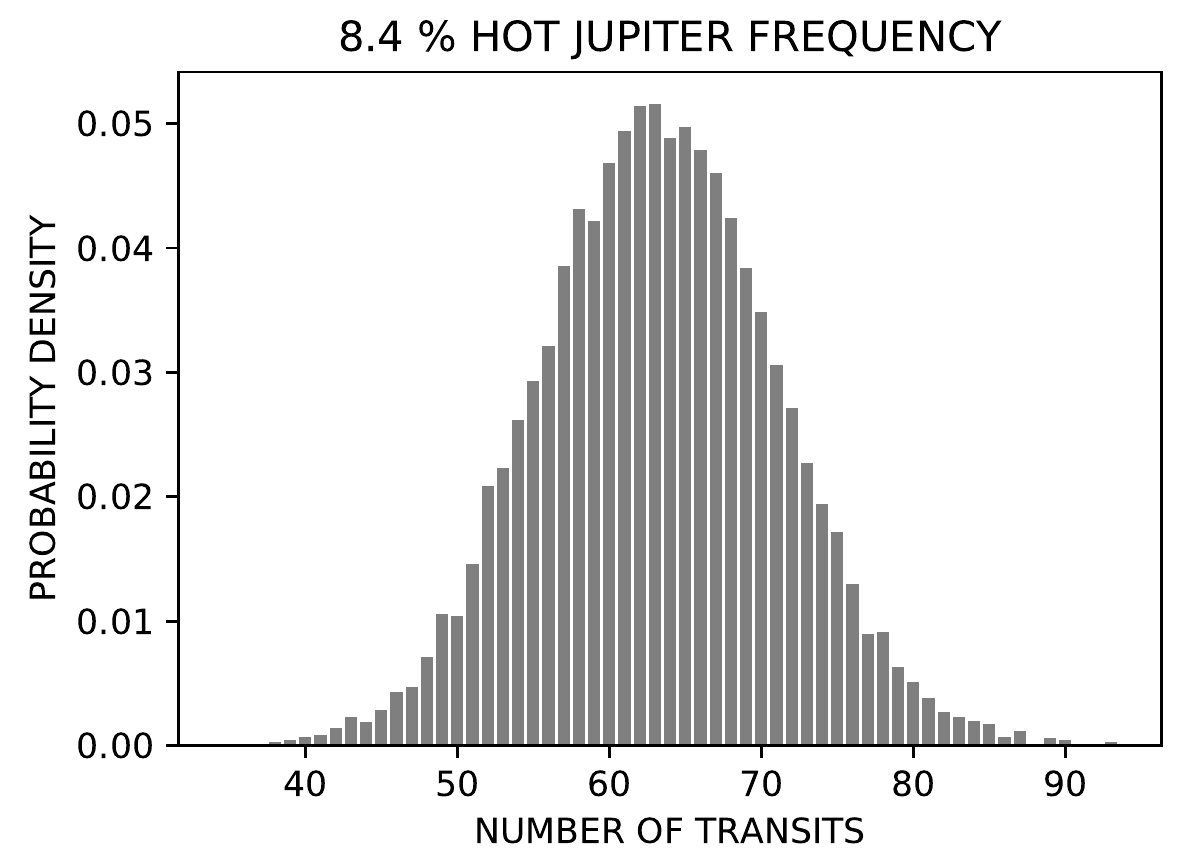}
\end{minipage}
 \caption{Transiting objects expected for different planet frequencies with 2\,$\sigma$; 0.15 per cent 0-3 Transits (1\,$\sigma$: 0-2 transits), 1.2 per cent 4-15 Transits (1\,$\sigma$: 6-12 transits), 0.75 per cent 2-10 Transits (1\,$\sigma$: 3-8 transits), 8.4 per cent 49-78 Transits (1\,$\sigma$: 55-71 transits)}
  \label{transit_number}
\end{figure}

To obtain a range of transits we would expect for different planet frequencies, we take the 2\,$\sigma$-range of the probability densities (or more precisely the 95 per cent values around the median value) -- see Fig.~\ref{transit_number}. For a frequency of 0.15 per cent as predicted by \cite{stephan2018} we therefore expect 0-3 transits and for a frequency of 1.2 per cent we expect 4-15 transits. Therefore, the lower frequency is much more likely. 

This raises the question whether we can find an upper limit for the giant planet frequency around \emph{Kepler} A-type stars. To obtain an estimate of the upper limit we increase the planet frequency of our simulation until the expected number of transits is at least 2. This is the case for a planet frequency of about 0.75 per cent (2-10 transits). This is a rather conservative upper limit considering that 1\,$\sigma$ or 68 per cent of all values already lead to an estimate of 3-8 transits.

If all 'possible planets' were indeed planets (planet frequency of 8.4 per cent) we would expect 49-78 transiting planets.

\section{Discussion}
\label{sectVI}
Combining the results from above a very high hot Jupiter frequency of 8.4 per cent (as in \cite{balona2014}) seems very unlikely for several reasons:
\begin{itemize}
 \item a transiting hot Jupiter shows up in the Lomb-Scargle periodogram of a light curve -- it therefore should be easily detected,
 \item the \emph{Kepler} team and pipeline must have missed around 130 transiting objects,
 \item we derive an upper limit on the hot Jupiter occurrence rate from \emph{Kepler} light curves of 0.75 per cent,
 \item the CoRoT survey found evidence that the number of hot Jupiters is similar for A-type and G-type stars and
 \item an upper limit on the occurrence rate of 4.5 per cent was determined on a sample of A-type stars observed with HARPS and SOPHIE.
\end{itemize}

In this section we explain our reasoning in more detail.

~

\begin{figure}
\includegraphics[width=\linewidth]{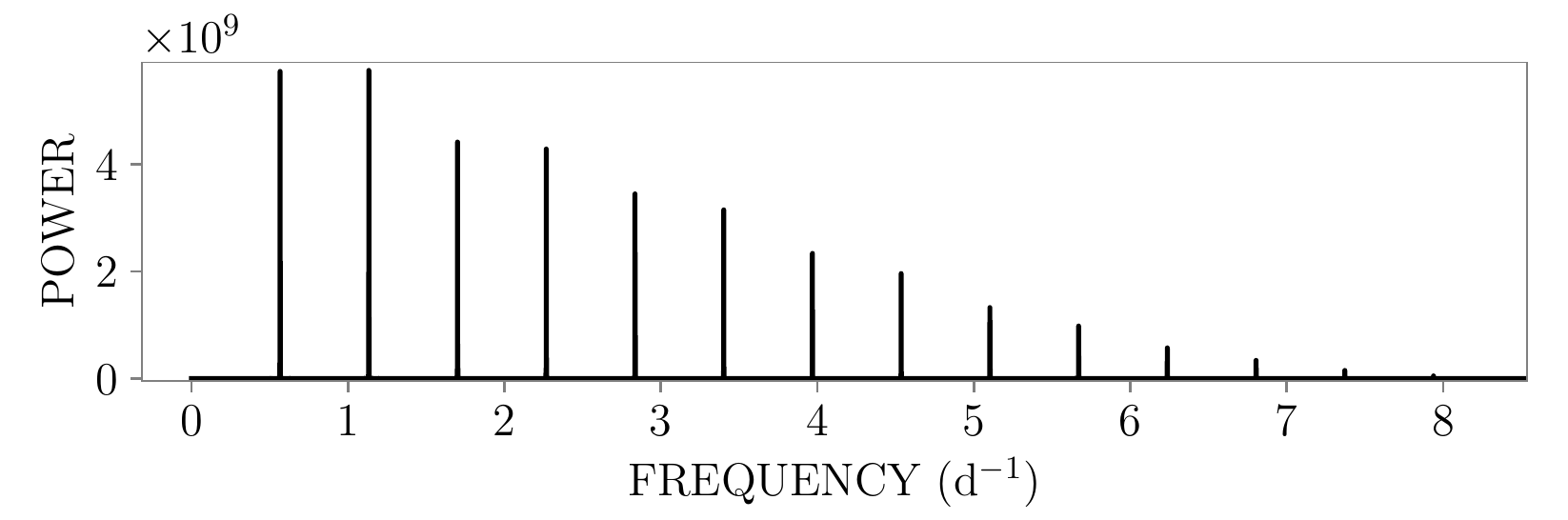}
 \caption{Periodogram of Kepler-13\,A\,b}
  \label{kepler13}
\end{figure}

Kepler-13\,A\,b is a Jupiter size planet that was discovered by \cite{shporer2011}. In Fig.~\ref{kepler13} we show the periodogram of its light curve. It becomes obvious that the rotation period of the star could not be retrieved without first subtracting the transits. This could be one reason why Kepler-13\,A\,b does not show up in the sample of \cite{balona2013}. 

We double checked this assumption with our transit blind tests (see Fig.~\ref{blindtest}) and can confirm that the transiting signal becomes the dominant feature in the periodogram for all hot Jupiters. In the 2013 sample there are two \emph{Kepler} planets (Kepler-340\,b\&c and Kepler-1115\,b). In those two examples the rotation period is very different from the candidate period such that the rotational peak can still be retrieved.

This means that the selection pattern used to describe the A-star activity in periodograms would exclude all transiting hot Jupiters. Nevertheless, if all 166 planets actually exist and all of them were found in \cite{balona2014} this means that, as a maximum, exactly the 66 per cent non-transiting planets were selected. The planet frequency would therefore not only be as high as 8.4 per cent but as high as 12.6 per cent. This means that the remaining around 1800 stars should show around 130 transiting hot Jupiters. The fact that by the \emph{Kepler} pipeline itself only one example of a transiting hot Jupiter was found makes it very improbable that 129 transiting planets were overlooked until now. This scenario is unlikely given that Kepler-13\,A\,b was easily detected.

In the CoRoT survey, \cite{guenther2016} found 9 candidates for planets of A-type stars. Three of them are binaries, one is a brown dwarf. Using AO-imaging and NIR spectroscopy they could exclude that they are false-positives. Using spectroscopy obtained with UVES they obtained upper limits of their masses which are all in the planetary regime. Assuming that these candidates are hot Jupiters, the number of hot Jupiters is the same for A-type as for G-type stars \citep{guenther2016}. With an occurrence rate of 8.4 per cent instead of $\approx$ 1.2 per cent the frequency of hot Jupiters were about 7 times larger for A-type stars than for G-stars. Consequently, they would have found 7 times as many candidates, because the photometric sensitivity of CoRoT was more than sufficient to detect hot Jupiters of A-type stars. The CoRoT survey thus shows that the frequency of hot Jupiters of A-type stars is about the same, or even less than that of G-stars. The CoRoT survey thus also excluded that the frequency of hot Jupiters of A-type stars is as high as 8.4 per cent.

Although RV-surveys of A-type stars are very challenging, \cite{borgniet2019} calculated the occurrence rate of close-in BD and giant planets from their sample with HARPS and SOPHIE. They concluded that the giant planet occurrence rate is lower than 4.5 per cent. In this paper we show that \emph{Kepler} retrieves an upper limit on the close-in hot Jupiter occurrence rate of 0.75 per cent. Combining the absence of transits in the \emph{Kepler} sample of A-type stars and our upper limits a very high hot Jupiter frequency of 8.4 per cent seems very unlikely. 

When compared to the theoretical predictions of the hot Jupiter frequency around main-sequence A-type stars our results are more consistent with the lower end of predictions like the ones from \cite{stephan2018}.

Nevertheless the question remains open what else could be the cause of the peculiar feature.

Although we find only upper limits in the range of 1.5-2.9\,M$_\text{Jup}$, we can disprove the hypothesis that a  close stellar companion is the source of the strange feature in the periodogram. A stellar companion would need to have an orbit of more than 30 years such that we could not find its trend in our two years of data. Due to Kepler's lower spacial resolution, a  stellar companion in an orbit of up to 1000 years could still be present in the same \emph{Kepler} pixel. 

\cite{saio2018} propose a different explanation for the broad and sharp feature in the periodogram. They suggest that the sharp feature could be a signature of spots on the star. The broad feature could then be explained by Rossby waves. They explain that there should be a link between the rotational frequency and that of the Rossby waves. Therefore they call those stars with the peculiar feature 'hump \& spike' stars. Nevertheless the question remains open why some of the rapidly rotating A-type stars show this kind of oscillations and some do not. In fact, in 81 per cent of the original \cite{balona2013} sample this kind of oscillations are not triggered.  

The hypothesis of \cite{saio2018} could be tested by taking a time series of spectra of the brightest targets of the sample. If the line distortions match the two periods causing the broad and the sharp peak this could indicate the presence of spots and oscillations. An exoplanet on the other hand does not cause line distortions that match its orbital period. This test was beyond the scope of this work and still needs to be conducted in the future.

Another explanation according to \cite{sikora2018} could be that the sharp peaks originate from inhomogeneities near the surface and the broad peaks in a region near a convective-radiative boundary, but their studies are still ongoing.  

\section{Conclusions}
\label{sectVII}

Aim of this work is to compare theories of planet formation and evolution with observations. Jupiter-sized planets around intermediate-mass main-sequence stars are still very rare. Theories predict either a high frequency of close-in Jupiter-like planets around these stars (>\,1.2 per cent) or a very low frequency (0.15 per cent).

\cite{balona2014} found peculiar features in light curves of 166 \emph{Kepler} stars. One possible explanation for this feature is the presence of close-in Jupiter sized companions. If these features are in fact the signature of close-in, massive planets it would mean that the frequency of such planets is a as high as 8 per cent. This is significantly higher than theories predict.

With observations from our two 2-m telescopes in Ond\v{r}ejov and Tautenburg we present upper limits of possible planetary-companion masses in the range of 3.8\,M$_\text{Jup}$-7.3\,M$_\text{Jup}$ for six of the stars in question. This excludes close-in stellar companions as well as planets that are heavier than 7.3\,M$_\text{Jup}$. 

We confirm that none of the planet candidates is transiting although statistically there should be around 80 transits. This makes the hypothesis of the 166 planets very unlikely. 

From the \emph{Kepler} sample of A-type stars there is only one detection of a hot Jupiter. As we have shown, transiting planets  with the size of Jupiter can easily be detected in the \emph{Kepler} data, even if they orbit A-type stars.

We derive that there is an upper limit of hot Jupiters of about 0.75 per cent. This is most consistent with the lower end of theoretical predictions. 

This is evidence for a lack of hot Jupiters around intermediate-mass main-sequence stars.

\section*{Acknowledgements}

This work was generously supported by the Th\"uringer Ministerium f\"ur Wirtschaft, Wissenschaft und Digitale Gesellschaft and the Deutsche Forschungsgemeinschaft (DFG) under the project GU 464/20-1.

MS acknowledges the Postdoc@MUNI project CZ.02.2.69/0.0/0.0/16-027/0008360.

JK and SG acknowledge support by Deutsche Forschungsgemeinschaft (DFG) grant PA525/18-1 within the DFG Schwerpunkt SPP 1992, Exploring the Diversity of Extra-solar Planets.

This paper includes data collected by the \emph{Kepler} mission.  Funding for the \emph{Kepler} mission is provided by the NASA Science Mission directorate.  

This research has made use of the SIMBAD database, operated at CDS, Strasbourg, France, and of data from the European Space Agency (ESA) mission Gaia ({https://www.cosmos.esa.int/gaia}), processed by the Gaia Data Processing and Analysis Consortium (DPAC, {https://www.cosmos.esa.int/web/gaia/dpac/consortium}).




\bibliographystyle{mnras}
\bibliography{literature.bib} 




\appendix

\section{Radial Velocities}

\begin{figure}
\includegraphics[width=\linewidth]{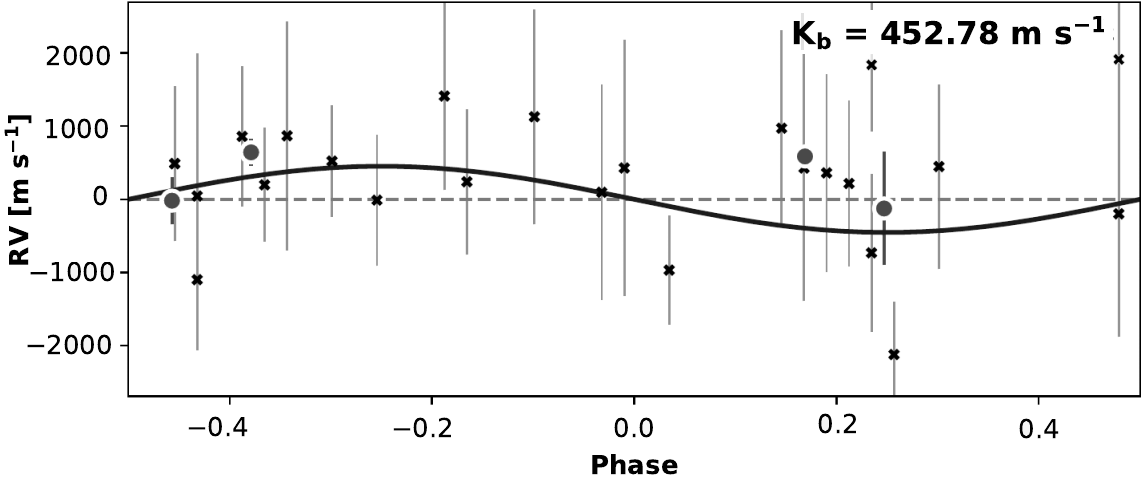}
 \caption{RV curve of KIC 3766112}
  \label{balona2}
\end{figure}

\begin{figure}
\includegraphics[width=\linewidth]{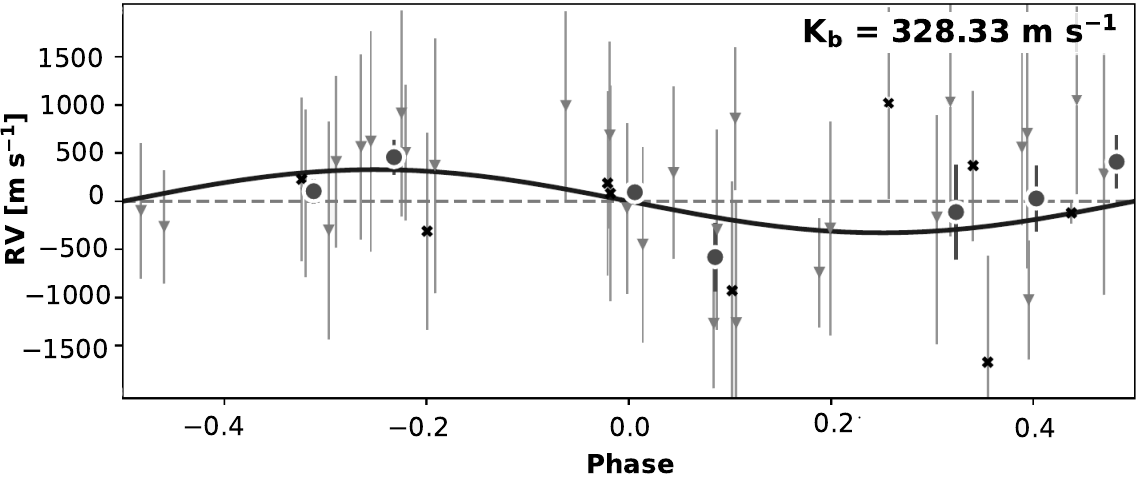}
 \caption{RV curve of KIC 4944828}
  \label{balona3}
\end{figure}

\begin{figure}
\includegraphics[width=\linewidth]{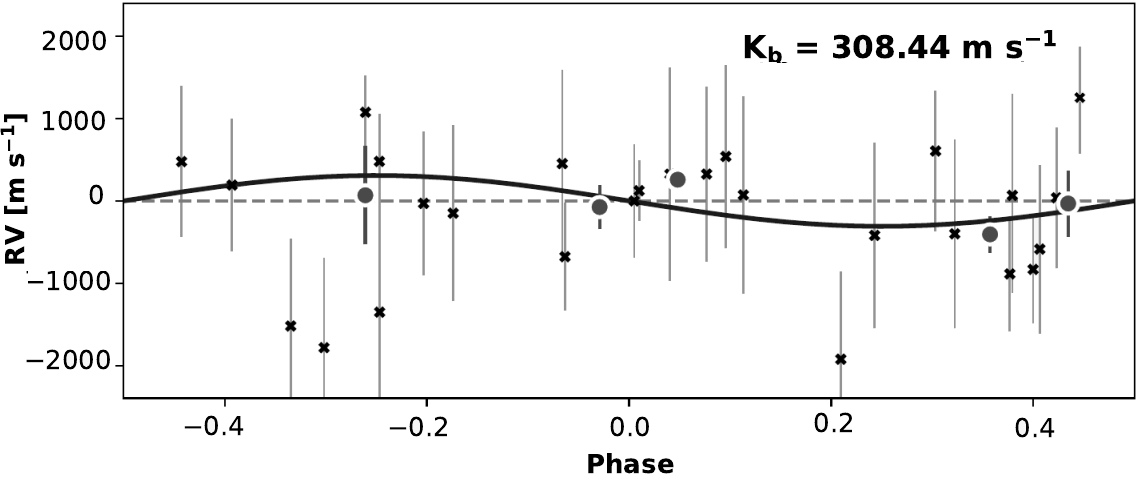}
 \caption{RV curve of KIC 7352016}
  \label{balona4}
\end{figure}

\begin{figure}
\includegraphics[width=\linewidth]{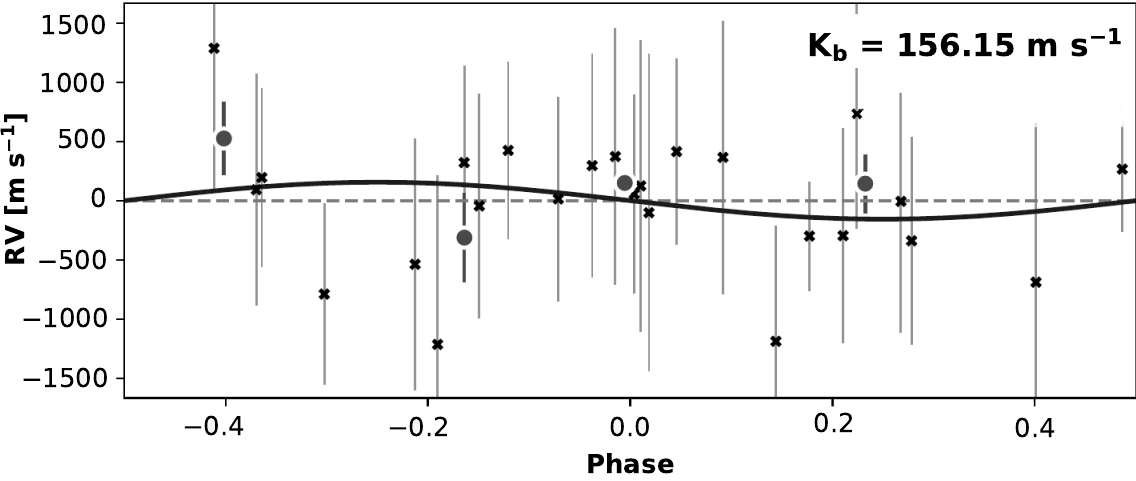}
 \caption{RV curve of KIC 7777535}
  \label{balona5}
\end{figure}

\begin{figure}
\includegraphics[width=\linewidth]{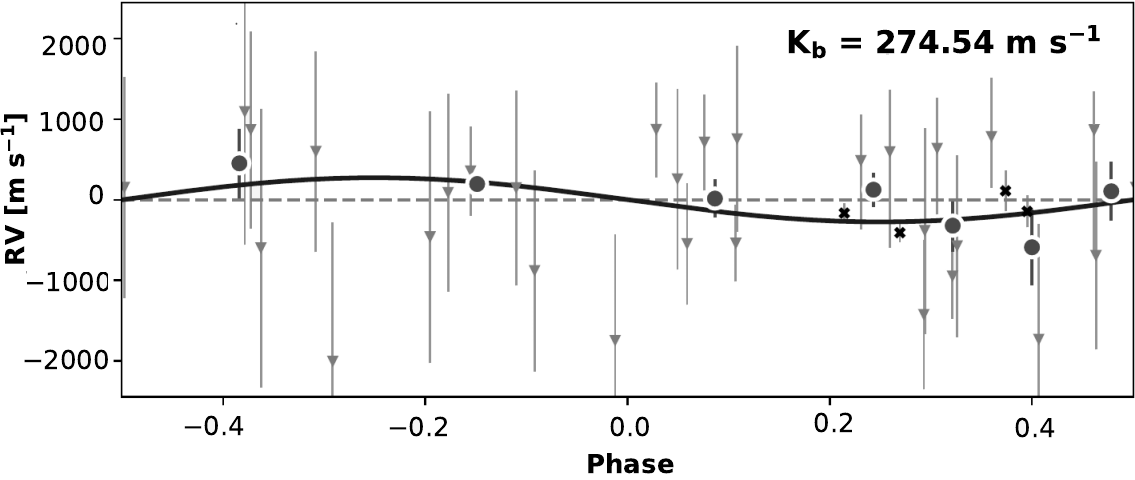}
 \caption{RV curve of KIC 9453452}
  \label{balona6}
\end{figure}

\begin{table}
  \begin{tabular}{llllll}
  \hline
  ~ &  \multicolumn{4}{c}{MASCARA-1-b Tautenburg data}  & ~ \\ 
    \hline
time & rv  & rv err  & time & rv  & rv err \\
(bjd) & ($\frac{\text{m}}{\text{s}}$) & ($\frac{\text{m}}{\text{s}}$) & (bjd) & ($\frac{\text{m}}{\text{s}}$) & ($\frac{\text{m}}{\text{s}}$)  \\
+2450000 & ~ & ~ & +2450000 & ~ & ~  \\
\hline
7966.36&2103&874&7980.35&2402&303\\
7966.39&2101&1027&7980.38&2393&1133\\
7966.43&1931&608&7980.4&1766&898\\
7966.46&1349&382&7980.42&1071&468\\
7966.47&1395&493&7980.46&1539&1061\\
7966.49&1735&820&7979.53&1244&859\\
7971.39&1494&1223&7980.55&983&694\\
7971.4&1833&591&7980.57&401&1265\\
7971.42&1363&484&7980.59&761&808\\
7971.43&1404&404&7995.33&2291&455\\
7971.45&2559&538&7995.35&1460&882\\
7971.46&2393&1360&7995.37&1389&338\\
7971.48&1185&863&7995.4&867&470\\
7971.49&1802&565&7995.42&1637&770\\
7971.53&539&704&7995.43&1225&566\\
7971.54&2554&758&7995.44&102&1260\\
7971.55&820&1189&7995.46&834&1219\\
7971.57&1370&1163&7995.47&953&851\\
7971.58&2285&1002&7995.49&628&633\\
7971.56&1885&465&7995.5&-309&662\\
7972.45&873&334&7995.52&611&1003\\
7972.47&667&869&7995.53&309&916\\
7972.48&978&799&7995.54&624&505\\
7972.49&1345&267&7995.57&1281&1758\\
7971.51&1046&683&7995.59&734&1520\\
7971.52&2120&836&7996.37&-418&2582\\
7971.54&1119&584&7996.39&-3&2173\\
7971.55&1709&945&7996.41&-54&2306\\
7971.56&1412&443&7998.34&670&489\\
7971.58&1328&536&7998.36&821&754\\
7971.55&1253&1104&7998.38&768&1266\\
7973.36&1242&539&7998.45&1397&692\\
7973.38&1594&864&7998.47&825&1287\\
7973.4&2265&780&7998.49&1159&1032\\
7973.42&1389&1125&7998.51&1612&989\\
7973.44&1725&580&7998.54&621&1372\\
7973.47&2027&510&7999.36&1129&1874\\
7973.49&2488&303&7999.51&1262&1040\\
7973.51&1728&425&7999.53&1603&1099\\
7973.53&1421&849&7999.55&1572&507\\
7973.55&1350&546&8000.41&315&1087\\
7973.57&1306&506&8000.43&626&834\\
7973.59&877&613&8000.45&840&762\\
7979.34&2343&1029&8000.47&937&828\\
7979.37&2170&1046&8000.49&508&437\\
7979.39&2065&844&8000.51&197&592\\
7979.41&2152&682&8000.53&891&432\\
7979.43&1557&1037&8000.55&1091&426\\
7979.45&2510&512&8000.57&1388&1534\\
7979.47&1102&270&8001.39&945&514\\
7979.49&1376&1076&8004.37&1454&643\\
7979.51&1806&337&8007.41&251&1559\\
7979.53&1706&353&8008.49&513&1610\\
7979.56&1521&736&8012.46&979&580\\
7979.58&1812&892&8013.4&549&952\\
7979.56&1268&1140&8014.3&1479&592\\
7980.33&2020&421&~&~&~\\
  \end{tabular}
    \caption{Barycentric Julian dates at mean exposure and the radial velocities determined from cross-correlation.} 
\end{table}

\begin{table}
\centering
  \begin{tabular}{lll}
  \hline
\multicolumn{3}{c}{MASCARA-1-b Ond\v{r}ejov data}  \\ 
    \hline
time & rv  & rv err  \\
(bjd) & ($\frac{\text{m}}{\text{s}}$) & ($\frac{\text{m}}{\text{s}}$)  \\
+2450000 & ~ & ~   \\
\hline
8313.41&791&1156\\
8313.48&5211&1054\\
8313.51&2453&2798\\
8313.54&2959&3167\\
8313.59&6032&3331\\
8314.56&2375&364\\
8314.41&1465&1712\\
8314.43&1412&1729\\
8334.4&2874&473\\
  \end{tabular}
    \caption{Barycentric Julian dates at mean exposure and the radial velocities determined from cross-correlation.} 
\end{table}

\begin{table}
  \begin{tabular}{llllll}
\hline
\multicolumn{3}{c}{KIC\,3766112} & \multicolumn{3}{c}{KIC\,4944828}  \\ 
    \hline
time & rv  & rv err  & time & rv  & rv err \\
(bjd) & ($\frac{\text{m}}{\text{s}}$) & ($\frac{\text{m}}{\text{s}}$) & (bjd) & ($\frac{\text{m}}{\text{s}}$) & ($\frac{\text{m}}{\text{s}}$)  \\
+2450000 & ~ & ~ & +2450000 & ~ & ~  \\
\hline
~ & ~ & \multicolumn{2}{c}{Tautenburg} & ~ & ~ \\
7557.49&-469&1402&7557.51&942&982\\
7558.46&74&1080&7558.51&-499&1016\\
7562.48&-1348&569&7562.51&970&1384\\
7563.49&-909&1604&7563.51&653&1396\\
7564.48&-976&1159&7564.51&232&1254\\
7566.51&-815&316&7585.49&246&897\\
7592.39&-1121&895&7585.53&-344&1039\\
7585.47&-1098&711&7588.52&-217&1315\\
7588.5&-1292&1818&7625.41&-122&887\\
7625.39&-848&801&7625.41&-122&887\\
7625.39&-848&801&7880.47&1000&976\\
7883.51&-366&1144&7883.53&515&962\\
7884.48&-889&1219&7884.5&454&704\\
7889.51&576&2206&7889.45&810&744\\
7911.45&798&989&7911.42&572&1146\\
7918.5&-209&1294&7918.53&-1075&620\\
7924.47&-910&1675&7923.47&360&892\\
7940.42&-477&656&7923.53&864&1073\\
7944.45&-2435&666&7924.49&320&1326\\
8001.46&-3458&207&7940.45&636&972\\
8008.38&-1139&348&7944.36&-790&569\\
8009.43&-1241&1297&7944.37&-332&1109\\
8012.36&-1535&1532&8001.53&-351&1132\\
8013.51&-2304&263&8007.46&-1321&675\\
8014.5&-2067&829&8008.41&-1314&1270\\
~&~&~&8012.39&510&805\\
~&~&~&8012.51&-149&705\\
~&~&~&8014.39&-315&590\\
~&~&~&8014.52&33&871\\
~ & ~ & \multicolumn{2}{c}{Ond\v{r}ejov} & ~ & ~ \\
~&~&~&7884.52&-279&1023\\
~&~&~&7892.47&-1641&1105\\
~&~&~&7905.39&1051&994\\
~&~&~&7926.51&113&1120\\
~&~&~&7929.41&-896&1135\\
~&~&~&7935.52&260&849\\
~&~&~&7946.36&402&787\\
~&~&~&8314.48&-89&111\\
~&~&~&8334.5&220&959\\
\hline
  \end{tabular}
    \caption{Part 1: Barycentric Julian dates at mean exposure and the radial velocities determined from cross-correlation.} 
\end{table}

\begin{table}
  \begin{tabular}{llllll}
\hline
\multicolumn{3}{c}{KIC\,7352016} & \multicolumn{3}{c}{KIC\,7777435}  \\ 
    \hline
time & rv  & rv err  & time & rv  & rv err \\
(bjd) & ($\frac{\text{m}}{\text{s}}$) & ($\frac{\text{m}}{\text{s}}$) & (bjd) & ($\frac{\text{m}}{\text{s}}$) & ($\frac{\text{m}}{\text{s}}$)  \\
+2450000 & ~ & ~ & +2450000 & ~ & ~  \\
\hline
~ & ~ & \multicolumn{2}{c}{Tautenburg} & ~ & ~ \\
7557.42&-325&1115&7557.53&206&977\\
7558.42&-653&303&7558.52&479&1154\\
7562.41&-172&951&7562.53&238&1235\\
7563.41&-711&1164&7563.53&380&531\\
7564.38&-740&829&7564.53&410&943\\
7566.41&-585&775&7585.51&128&864\\
7589.43&-1664&666&7588.42&846&968\\
7585.40&-454&1278&7588.54&-574&1348\\
7588.45&-1176&1129&7625.43&69&951\\
7625.33&-778&657&7883.55&435&815\\
7625.52&-2699&1042&7884.52&108&1112\\
7880.49&-300&891&7889.48&1400&1186\\
7883.46&-298&538&7911.47&528&785\\
7884.43&-808&847&7919.44&-1099&1427\\
7884.56&-1454&618&7923.49&-423&1062\\
7914.43&-452&1040&7924.50&-225&877\\
7919.42&475&643&7940.49&538&747\\
7923.41&298&392&7944.39&309&759\\
7924.42&-926&1048&8001.55&170&842\\
7940.47&-239&1095&8007.44&-674&765\\
7944.47&-1608&617&8008.42&-1073&975\\
8001.49&-2128&1345&8012.40&12&1339\\
8008.33&-706&1178&8012.53&-182&910\\
8009.38&-1196&1111&8014.41&486&1082\\
8012.32&-1363&1003&8014.54&-185&463\\
8013.49&-2296&1035&~&~&~\\
8014.45&-2559&1071&~&~&~\\
\hline
  \end{tabular}
    \caption{Part 2: Barycentric Julian dates at mean exposure and the radial velocities determined from cross-correlation.} 
\end{table}

\begin{table}
  \begin{tabular}{llllll}
\hline
\multicolumn{3}{c}{KIC\,9222948} & \multicolumn{3}{c}{KIC\,9453452}  \\ 
    \hline
time & rv  & rv err  & time & rv  & rv err \\
(bjd) & ($\frac{\text{m}}{\text{s}}$) & ($\frac{\text{m}}{\text{s}}$) & (bjd) & ($\frac{\text{m}}{\text{s}}$) & ($\frac{\text{m}}{\text{s}}$)  \\
+2450000 & ~ & ~ & +2450000 & ~ & ~  \\
\hline
~ & ~ & \multicolumn{2}{c}{Tautenburg} & ~ & ~ \\
7557.44&-668&893&7557.40&-13&1176\\
7558.44&-316&737&7558.39&-461&1197\\
7562.46&-450&771&7562.38&256&1553\\
7563.46&-272&836&7563.38&148&1144\\
7564.46&-968&1265&7564.41&-1069&1557\\
7566.49&-1014&1236&7566.38&-350&1111\\
7590.43&-1583&1348&7568.39&174&609\\
7585.42&-1530&1150&7585.35&-2035&911\\
7588.47&-1763&956&7585.37&-1182&1117\\
7625.36&-1614&825&7625.31&-1147&442\\
7625.36&-1614&825&7625.31&-1147&442\\
7880.52&-927&1419&7880.44&28&800\\
7883.48&-734&1257&7883.43&-124&832\\
7884.45&-1416&1451&7884.41&-251&527\\
7889.43&-827&1288&7884.54&-1155&736\\
7911.49&823&787&7889.54&-992&1270\\
7912.46&485&627&7912.48&105&571\\
7919.46&82&1328&7919.49&484&1641\\
7923.43&0&1196&7923.38&263&565\\
7924.44&-966&1310&7924.39&-9&1235\\
7941.44&-511&781&7939.53&262&1212\\
7944.49&-366&714&7941.47&-519&1215\\
8001.57&-1764&1428&7942.38&-1558&496\\
8008.36&-2940&7302&7942.49&-456&1362\\
8009.41&-1095&1583&8001.51&-2615&1720\\
8012.34&-809&1645&8008.31&-1489&1240\\
8014.37&2635&5670&8009.36&-1207&1720\\
8014.47&-1946&1670&8012.29&-1302&1153\\
~&~&~&8013.47&-2342&1423\\
~&~&~&8014.43&-2357&1313\\
~ & ~ & \multicolumn{2}{c}{Ond\v{r}ejov} & ~ & ~ \\
7892.42&-366&415&7891.41&251&254\\
7929.45&476&993&7928.46&-3&199\\
7948.37&405&69&7948.42&-268&115\\
7995.52&-1512&606&8322.40&-25&126\\
8334.58&-332&167&~&~&~\\
\hline
  \end{tabular}
    \caption{Part 3: Barycentric Julian dates at mean exposure and the radial velocities determined from cross-correlation.} 
\end{table}


\bsp	
\label{lastpage}
\end{document}